# Evolutionary Neural Architecture Search for Retinal Vessel Segmentation

Zhun Fan[1,2], *Senior Member, IEEE,* Jiahong Wei[1,2], Guijie Zhu[1,2], Jiajie Mo[1,2], and Wenji Li[1,2]

[1]Department of Electronic and Information Engineering, Shantou University, Shantou 515063, China
[2]Key Lab of Digital Signal and Image Processing of Guangdong Province, Shantou University, Shantou 515063, China
zfan@stu.edu.cn; 19jhwei@stu.edu.cn; 16gjzhu@stu.edu.cn; jiajiemo@outlook.com; liwj@stu.edu.cn

**Abstract:**

The accurate retinal vessel segmentation (RVS) is of great significance to assist doctors in the diagnosis of ophthalmology diseases and other systemic diseases. Manually designing a valid neural network architecture for retinal vessel segmentation requires high expertise and a large workload. In order to improve the performance of vessel segmentation and reduce the workload of manually designing neural-network, we propose novel approach which applies neural architecture search (NAS) to optimize an encoder-decoder architecture for retinal vessel segmentation. A modified evolutionary algorithm is used to evolve the architectures of encoder-decoder framework with limited computing resources. The evolved model obtained by the proposed approach achieves top performance among all compared methods on the three datasets, namely DRIVE, STARE and CHASE_DB1, but with much fewer parameters. Moreover, the results of cross-training show that the evolved model is with considerable scalability, which indicates a great potential for clinical disease diagnosis.

## 1. Introduction

Retinal vessel segmentation has played an important role in the field of medical image processing, because the pathological changes of retinal blood vessels can reflect either ophthalmology diseases or other systemic diseases, such as high blood pressure, diabetes, arteriolosclerosis, and so on. At present, ophthalmologists and other doctors consider the fundus examination as a routine clinical examination [56]. Moreover, the fundus vascular system is the only human blood vascular system that can be observed in vivo [57][58]. Through the observation of the fundus vascular system, we can diagnose and track many diseases [57]. Retinal vessel segmentation is a prerequisite step for quantitative analysis of fundus images. Through the retinal vessel segmentation, the relevant morphological information of retinal vascular tree (such as the width, length and curvature of blood vessels, etc.) can be obtained [31], which can be applied in biometrics [59][60]. So accurate segmentation of retinal vessels is of great significance.

  The structure of the retinal vasculature tree is very complicated, with lots of interconnected blood vessels, including many tiny blood vessels. Actually, the difference between the vascular region and the background is subtle, and the fundus images are susceptible to noise and uneven illumination. As a result, it is very challenging to segment the retinal vascular trees from fundus images. Although many manually designed neural network architectures for retinal vessel segmentation have been proposed, they have some limitations. The existing neural network models are still difficult to capture vascular trees under complicated situations of the fundus images. Therefore, it is necessary to design a neural network model with carefully optimized architecture to extract the features of the complicated vascular tree more accurately in these situations.

In this paper, we apply neural architecture search (NAS) with an evolutionary algorithm as optimization method to retinal vessel segmentation (RVS). In order to improve the performance of RVS–with an optimized neural network architecture, we propose a specific search space based on encoder-decoder framework inspired by U-Net[2]. For macro-architecture search, we also adopt an encoding method with fixed length and non-redundant binary code to represent the neural network architecture and use modified evolutionary strategy (ES) to search efficiently under the large search space. ES can evolve with smaller populations in situations of limited computing resources. As shown in Figure 1, during the architecture evolution, we adopt standard operations of evolutionary strategy (ES) (e.g. selection and mutation) to produce more competitive neural network architectures.

Our specific contributions are as follows:
(1) It is the first time that NAS is applied to retinal vessel segmentation.
(2) We propose a specific search space based on encoder-decoder framework, and use modified evolutionary strategy to automatically optimize the neural network architecture.
(3) The searched model achieves the top performance on three public available datasets, DRIVE [13], STARE [14], CHASE_DB1 [39]. Meanwhile, cross-training between these three datasets verify the robustness and scalability of the evolved model.

The remainder of this paper is structured as follows: Section 2 reviews related work with a focus on retinal vessel segmentation and neural architecture search. Section 3 introduces the proposed method. Section 4 presents the evaluation metrics, loss function and datasets. Section 5 describes the result of architecture evolution and experimental results. Finally, we conclude this work in Section 6.

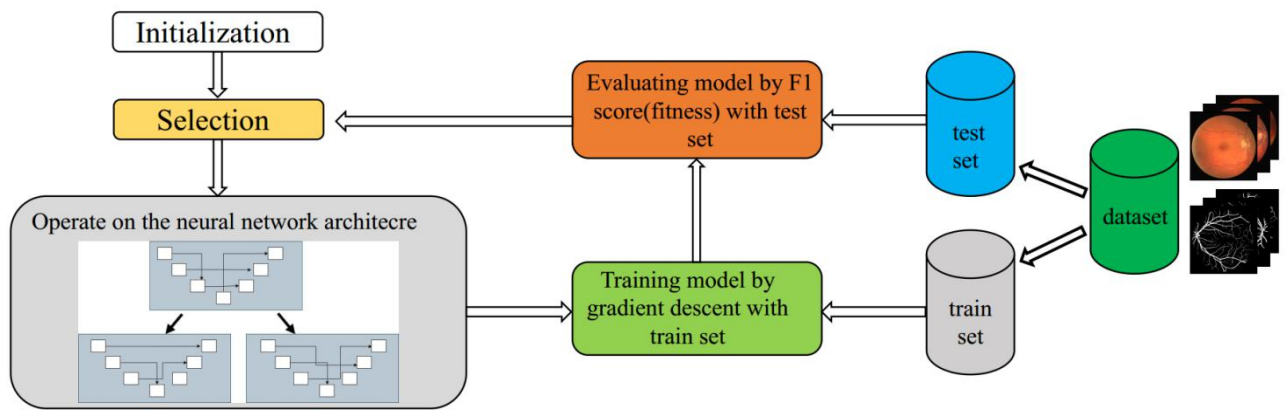

Figure 1. Overview of our method. Our approach is an architecture search process based on evolutionary strategy. the model is trained on a retina vessel dataset and assigned F1-score of the trained model as the fitness. The evolution strategy will search for better architectures.

## 2. Related work

### 2.1 Retinal vessel segmentation

Since FCN[1] and U-Net[2] were proposed, image segmentation methods based on full convolutional neural networks have become mainstream because of their excellent effects. Retinal vessel segmentation is a subclass of image segmentation. Recently, the new state-of-the-art methods [55][54][31][30][29][40] for retinal vessel segmentation are dominated by neural network models, especially variants of U-Net.

[55] adopts a joint-loss to train the U-Net model. Two branches are responsible for pixel-wise loss and segment-level loss, respectively. The joint-loss can promote the model to balance the thick and thin vessels. [54] adds pretrained components of MobileNetV2 [61] as the encoder and introduces

novel contractive bottleneck bocks as the decoder, which achieves better performance and faster inference speed. [31] replaces traditional convolution with deformable convolution into U-Net to capture the miscellaneous morphology of vascular trees. [40] includes dense dilated convolutional block between the same stage encoder cell and decoder cell of the U-Net, and also uses a regularized walk algorithm to post-process model prediction. [29] designs a novel inception-residual block and introduces four supervision paths with different kernel sizes of convolution to utilize multi-scale features. The model in [30] has two encoders based on U-Net. One encoder path is for extracting spatial information and the other path is for extracting context information. In addition, a novel module is used to combine the information of the two paths.

## 2.2 Neural architecture search

Depending on the search method, NAS can be divided into three categories: reinforcement learning based, evolutionary algorithm based and differential architecture search. The method based on reinforcement learning [42][43][44] is to sample neural network architectures and use a controller to learn how to generate better architectures from continuous trial and error, with the performance of the models as reward feedback to the controller. The method based on evolutionary algorithms [11][12] is to perform operations on the neural network architecture (such as crossover and mutation) to generate off-springs, and to continuously adjust the neural network architectures from generation to generation according to the general principle of survival of the fittest, and finally obtain the optimized model. For differential neural architecture search [45][46], each operation of the cell is assigned a weight coefficient. The parameter weight of neural network and the weight of each operation are updated alternatively by gradient descent. The optimal model can be obtained by selecting the operation with the largest weight after convergence.

After successful application of NAS in image recognition, some researchers have also extended NAS to image segmentation [4] and object detection [47][48], including some works applying NAS to medical image segmentation. [49], [50] and [51] are mainly to optimize the hyperparameters and operations of each layer of the neural network. [52] and [53] belong to cell-based method that optimizes the structure and operations inside the cell based on U-Net. However, it is noteworthy to point out that there is no work of applying NAS to retinal vessel segmentation yet.

## 2.3 Evolutionary strategy

In the traditional evolutionary strategy [62], each individual is represented by decision variable X and standard deviation σ. They all contain $n$ components, as follows:

$$[X, \sigma] = [(x_1, x_2, ..., x_i, ..., x_n), (\sigma_1, \sigma_2, ..., \sigma_i, ..., \sigma_n)]$$

The relationship between X and σ is:

$$\sigma_i' = \sigma_i \cdot \exp\left(r' \cdot N(0,1) + r \cdot N_i(0,1)\right)$$

$$x_i' = x_i + \sigma_i \cdot N_i(0,1)$$

$(x_i, \sigma_i)$ is the i-th component of the parents. $(x_i', \sigma_i')$ is the i-th component of the off-springs. $N(0,1)$ is a random number subject to the standard normal distribution. $N_i(0,1)$ is a random number conforming to the standard normal distribution for the i-th component. $r'$ is global coefficient and r is local coefficient. It indicates that old individuals generate new individuals through Gaussian mutation.

In the evolutionary process, the evolutionary strategy consists of two main steps: (1) In each generation, change the decision variable X of each individual through mutation or recombination operation; (2) Keep the best individuals for the next generation and repeat the process.

Evolutionary strategy can be divided into two main forms: (μ,λ) and (μ + λ). Both (μ,λ) and (μ + λ) produce λ off-springs from μ parents. (μ,λ) evolutionary strategy ( λ > μ ) selects the best μ individuals from λ off-springs for next generation, and ignore the parents. Even though the parents might be better than the off-springs. However, the (μ + λ) evolutionary strategy compares parents and off-springs together, and selects the best μ individuals for next generation. (μ + λ) evolution strategy keeps elite individuals in the whole evolution process. The evolutionary strategy has a flexible population size and can evolve in a smaller population, which is conducive to the optimization of neural network under the situation of limited computing resources.

## 3. The proposed method

In this section, we will focus on how to use NAS for RVS and the optimization method of neural network architecture. To improve the effect of RVS by a novel neural network architecture, it is necessary to specify the search space in the NAS process. In addition, because the evolutionary algorithm is used to automatically design the architecture of the neural network, the representation of the neural network architecture for evolving is also very critical, which will be explained in detail in this section.

### 3.1 The search space

#### 3.1.1 The basic framework of the network architecture

Since FCN[1] and original U-Net[2] with skip connections added for better feature fusion, full convolutional neural networks with encoder-decoder structure are currently mainstream in image segmentation. Due to the outstanding performance of original U-Net and its transferability, the U-like neural network architectures (variations of original U-Net) are still a common choice in medical image segmentation. The architecture of the U-like neural network is composed of an encoder for down-sampling and a decoder for up-sampling. The encoder extracts image features of different scales, and the decoder recovers the extracted features in the encoder to the original image sizes and classify each pixel in the original image. With the above consideration, the basic framework of the neural network architecture in our method is also a U-like architecture with several different cells. It mainly contains one initial-conv cell, three encoder cells, and three decoder cells, which means three down-sampling and three up-sampling.

#### 3.1.2 Search space for the cells

In many NAS methods[52][53][46][45], cell-based micro-search methods are used, which is a search of operations inside the cell, and the searched neural network architecture is formed by stacking several of these cells with same structure and operations. Our method will search the internal operations of cells and skip connections between different cells, which is a macro-search method different from cell-based micro-search. Inspired by the [3], our method also includes several cells. Each cell would contain different operations. As shown in Table 1 and Figure 2, the search space of our method mainly includes the normalization methods, the activation functions, the up-sampling methods and the down-sampling methods, the shortcut connections inside the cells, and the skip connections with the preceding cells. At the same time, unlike original U-Net, which uses fixed skip connections for feature fusion between cells, our method searches skip connections between the cells for optimal feature fusion, resulting in an optimized network architecture. In order to mitigate computational complexity, each cell utilizes summation to fuse the features from skip connections, while original U-Net utilizes concatenation. In addition to skip connections between different cells, the search space also contains the operation of whether to use the shortcut connection inside the cell.

**Table 1. The proposed search spaces**

| Operation | Search space | Feasible cell |
|---|---|---|
| Activation function | ReLU[7], SELU[8] | All cells |
| Normalization | Batch normalization[9], Instance normalization[10] or no normalization | All cells |
| Down-sampling | Max pooling, Average pooling | Encoder cells |
| Up-sampling | Bilinear interpolation, Transpose convolution | Decoder cells |
| Shortcut connection | Yes / No | Encoder cells, Decoder cells |
| Skip connection | Connections with preceding cells | Encoder cells, Decoder cells |

We also add some commonly used activation functions, normalization methods and sampling methods to the search space, which is to optimize the settings of the network model. The number of convolution kernels in all cells is set to 128. When dimension mismatch occurs in connections (skip connections or shortcut connections), we use 1 × 1 convolution to adjust the channel number of feature map from connections or resize the feature map by nearest neighbor interpolation.

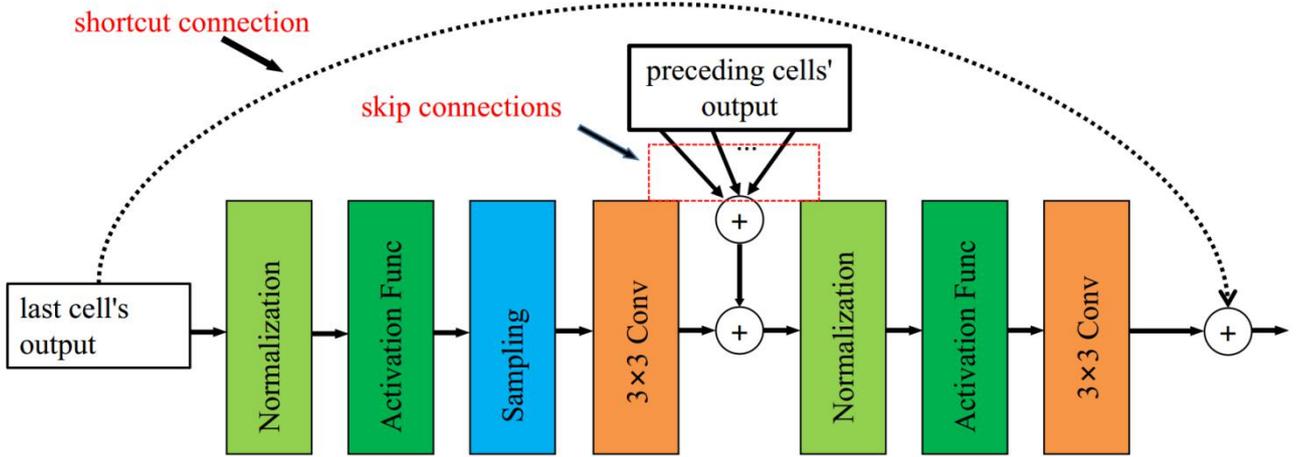

Figure 2. The proposed search space of the cells (There isn't shortcut connection, up-sampling and down-sampling in initial-conv cell. The difference between encoder cell and decoder cell is mainly reflected in up-sampling and down-sampling.)

### 3.2 Representation of the neural network

Similar to [11][12], our method also adopts the binary bit to encode the network architecture. In order to search the architecture more efficiently, the coding of the architecture should be as short as possible, and have a one-to-one correspondence between genotype and phenotype. As shown in Figure 3, we use a tuple ($S, N_1, N_2, A, SC, SK_1, ..., SK_i$) with binary values to represent each cell, and each bit represents the corresponding operation.

- $S$ stands for sampling method of the corresponding cell. It includes Max pooling or Average pooling for encoder cells, and bilinear interpolation or transpose convolution for decoder cells.
- $N_1$ indicates whether to use normalization.
- $N_2$ indicates which type of normalization, batch normalization or instance normalization.
- $A$ is a bit indicating which type of activation function is used, ReLU or SELU.

- *SC* indicates the in-cell shortcut connection.
- *SK$_i$* means whether to take a skip connection with i-th preceding cell.

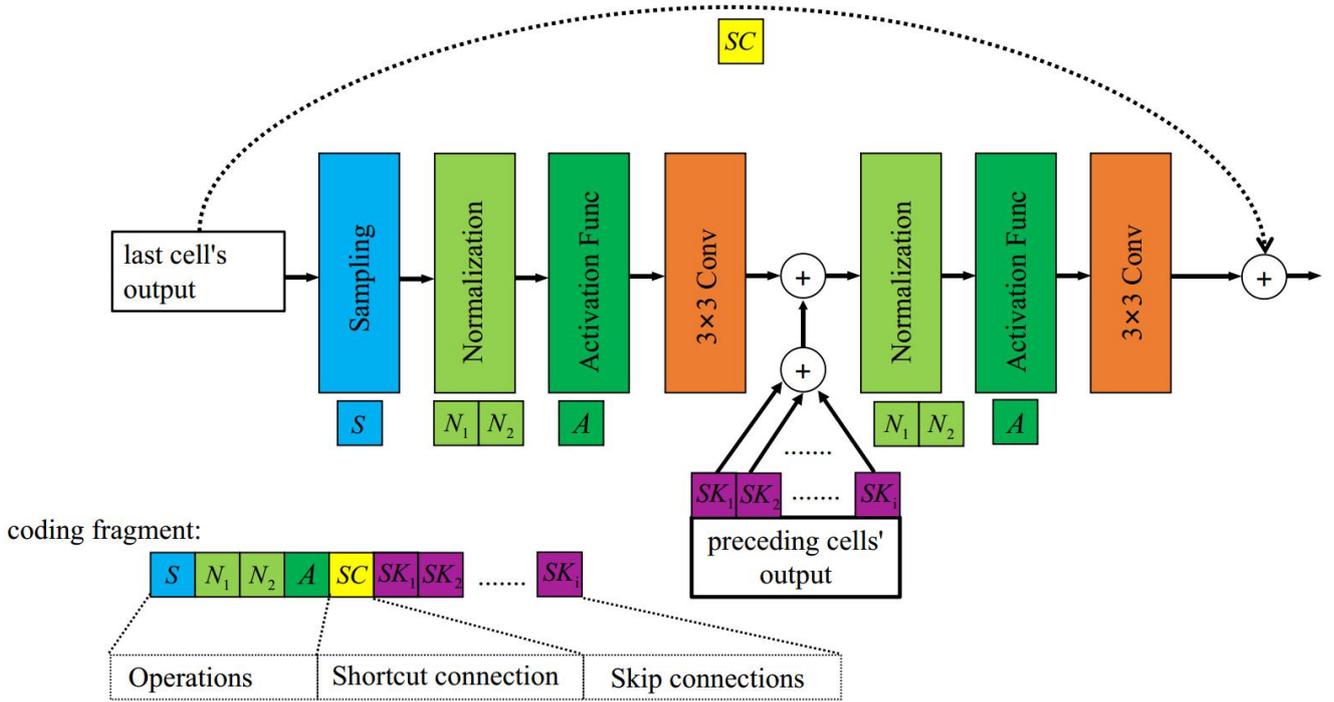

Figure 3. Example of genotype and phenotype. This is the cell encoding fragment and its corresponding phenotype. Different bits indicate different operation or connections. S, N, A, SC, SK represent sampling method, normalization method, activation function, shortcut connection, skip connection, respectively.

It should be noted that there are no *S, SC* and *SK* in the initial-conv cell. Often the rear cells will be with more bits of *SK*, since they can have more skip connections from preceding cells. The genotype of the whole neural network architecture is spliced by code fragments of different cells.

## 3.3 Optimization method

### 3.3.1 Modified (μ + λ) evolution strategy

The traditional evolutionary strategy uses real number encoding, and uses Gaussian mutation to generate more competitive individuals. Our method modifies the traditional (μ + λ) evolution strategy. The main changes are: (1) Changing the encoding mode, using the binary code encoding neural network models; (2) Changing the mutation mode, the binary code mutates with random flip of bits; (2) Introducing the binary code crossover operation.

In addition, in the process of architecture evolution, it takes lots of computing resources and computing time to evaluate the candidates. In order to balance the limited computing resources and the efficiency of evolution, our method gradually reduces the population size in stages, which is to reduce the number of candidates for evaluation while ensuring the search ability. As shown in Algorithm I, we divide the modified (μ + λ) evolution strategy into three stages. In the first stage (μ = 18, λ = 18), in order to ensure that the global search ability, our method evolves in a relatively large population. Before the start of the next two stages, we reduce the population size by keeping fewer elite individuals in the last generation of the previous stage. In the third stage (μ = 4, λ = 4), in order to enhance the local search ability, only mutation is used, and the probability of mutating per bit decrease from 0.1 to 0.05. We select the best individual from the last generation as the optimal solution.

**Algorithm I:** Neural network evolution

**Input:** Parent size $\mu$, off-spring size $\lambda$, number of generations in first stage $N_1$, number of generations in second stage $N_2$, number of generations in third stage $N_3$, probability of individual mutation $p_i$, mutation probability of each bit $p_b$.
**Output:** The best individual and its corresponding DNN model.
1    $P_1 \leftarrow$ Initialize population ($\mu = 18$, $dnn$);
2    evaluate ($P_1$, $dnn$);
3    **for** $g = 1$ to $N_1 + N_2 + N_3$ **do**
4       **if** $g <= N_1$ **then**
5          $O' \leftarrow$ crossover ($P_1$, $\lambda = 18$);
6          $O \leftarrow$ mutate ($O'$, $p_i = 0.4$, $p_b = 0.1$);
7          evaluate ($O$, $dnn$);
8          $P_1 \leftarrow$ select Best ($O + P_1$, $\lambda = 18$);
9       **else if** $N_1 < g <= N_2$ **then**
10        **if** $g = N_1 + 1$ **then**
11           $P_2 \leftarrow$ select Best ($P_1$, $\lambda = 8$);
12        $O' \leftarrow$ crossover ($P_2$, $\lambda = 8$);
13        $O \leftarrow$ mutate ($O'$, $p_i = 0.4$, $p_b = 0.1$);
14        evaluate ($O$, $dnn$);
15        $P_1 \leftarrow$ select Best ($O + P_2$, $\lambda = 8$);
16      **else if** $N_2 < g <= N_1 + N_2 + N_3$ **then**
17        **if** $g = N_2 + 1$ **then**
18           $P_3 \leftarrow$ select Best ($P_2$, $\lambda = 4$);
19        $O \leftarrow$ mutate ($P_3$, $p_i = 1$, $p_b = 0.05$);
20        evaluate ($O$, $dnn$);
21        $P_3 \leftarrow$ select Best ($O + P$, $\lambda = 4$);
22   **end**
23   Best $\leftarrow$ select Best ($P_3$, $\lambda = 1$);
24   **return** Best and its corresponding DNN model;

### 3.3.2 Population initialization

In [36], an initialization method called Rich Initialization is introduced, and has proven to work better than random initialization. Inspired by this, we also adopt a similar initialization method which generates a shortcut connection in all encoder and decoder cells of the network architecture represented by the initial P, and all encoder and decoder cells will be connected with all preceding cells via skip connections.

## 4. Material

### 4.1 Loss function

Focal loss[18] is proposed to cope with the imbalance of positive and negative samples. The loss function adopted in this work is given in Equation (1), where $y, \hat{y}, n, m$ denote ground truth, model prediction, n-th sample, and the total number of all samples, respectively.

$$L = - \sum_{n=1}^{m} \left( \alpha y_n (1 - \hat{y}_n)^\omega \log \hat{y}_n + (1 - \alpha)(1 - y_n) \hat{y}_n^\omega \log(1 - \hat{y}_n) \right) \qquad (1)$$

## 4.2 Datasets

In the work, we use three public datasets: DRIVE[13], CHASE_DB1[14], STARE[39]. An overview of these 3 publicly available datasets is provided in Table 2. Some examples of the datasets are also shown in Figure 4.

Table 2. Overview of the adopted datasets in this paper

| Dataset | quantity | Resolution | Train-test split |
| --- | --- | --- | --- |
| DRIVE | 40 | 565 × 584 | Official train-test split |
| STARE | 20 | 700 × 605 | Leave-one-out split |
| CHASE_DB1 | 28 | 999 × 960 | First 20 for train, last 8 for test |

The three public datasets contain the manual annotations of two experts, and we only take the annotations of the first expert as the ground truth. Unlike the patch-based methods, we use the original images without cropping as the input of the models. CHASE_DB1 and STARE do not have a predefined train-test split. In order to compare with other methods, we use the same train-test split as in [15][23][28].

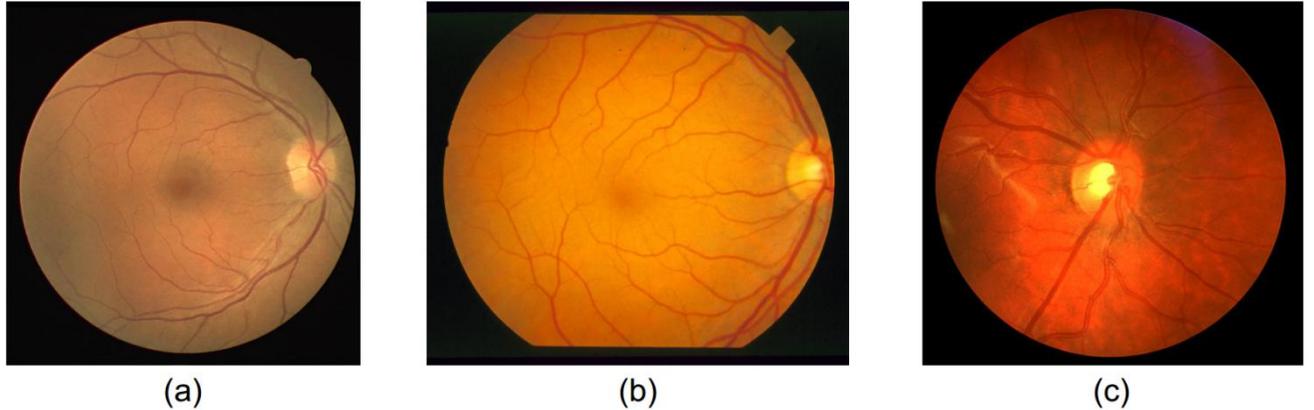

Figure 4. Examples of training images (H × W). (a): DRIVE (584 × 565); (b): STARE (700 × 605); (c): CHASE_DB1 (960 × 999)

## 4.3 Evaluation Metrics

RVS, a binary classification problem, is to predict whether the pixels of retina vessel images belong to vessel (positive) or non-vessel(negative). *TP, FP, TN, FN* represent true positive, false positive, true negative, false negative, respectively. As shown in Table 3, there are five metrics selected. These metrics are all based on *TP, FP, TN, FN*. In our work, the global threshold $\tau$ is set to 0.5 when calculating *TP, FP, TN, FN,* except for *AUROC* which is calculated by different thresholds.

Table 3. The evaluation metrics in our method

| Metric | Description |
| --- | --- |
| *ACC* (accuracy) | *ACC = (TP + TN) / (TP + TN + FP + FN)* |
| *SE* (sensitivity) | *SE = TP / (TP + FN)* |
| *SP* (specificity) | *SP = TN / (TN + FP)* |
| *F1-score*(F1) | *F1 = (2 ×TP) / (2 × TP + FP + FN)*      (2) |
| *AUROC* | Area Under the ROC curve. |

## 4.4 Objective Function for ES

RVS is not only a dense prediction problem, but also an imbalanced classification problem. In a fundus image, the non-vessel area is more than 90%. Therefore, when evaluating the model performance, more comprehensive metrics, such as *F1-score* and *AUROC*, need to be considered. In the process of evolving the optimal network architecture using modified ES, our work uses *F1-score* as fitness for this single-objective optimization problem, so the goal of optimization is to maximize Equation (2).

## 5. Experiments

### 5.1 Implement details

All experiments are performed on a GPU server, which has four NVIDIA TITAN Xp GPUs with 12GB memory each. During the evolution of the neural network architecture, each architecture in the evolution process needs to be evaluated. The evaluation method is to first train the model by gradient descent using training set. Then the test set is used to evaluate the performance of the model, and the result is assigned to the corresponding individual as its fitness. In the above training process, we use DRIVE dataset with a batch size of 2. At the same time, the optimizer we used is Lookahead[16] with Adam[17] as the base optimizer ($\beta_1$ = 0.9, $\beta_2$ = 0.999). The Lookahead optimizer is adopted default parameters, such as $\alpha$ = 0.5, $k$ = 6. Among them, the learning rate initialized during the training process is 0.001, and gradient is clipped with the L1 norm threshold of 0.1.

To deal with the data imbalance of RVS, we used Focal loss[18] (The two parameters, $w$ and $\gamma$, are set to 0.55 and 2.0, respectively). During the evolution of the neural network architecture, the pixels of the images are normalized to the range of [-1,1]. We train each neural network model for 100 epochs, and take the F1-score calculated on the test set as the fitness of the corresponding individual in the evolution process.

During the training process, F1-score is calculated for each epoch. If the F1-score does not change after 20 epochs, it is considered that the model has been fully optimized, and needs no further training to save computing resources.

### 5.2 Results of architecture evolution

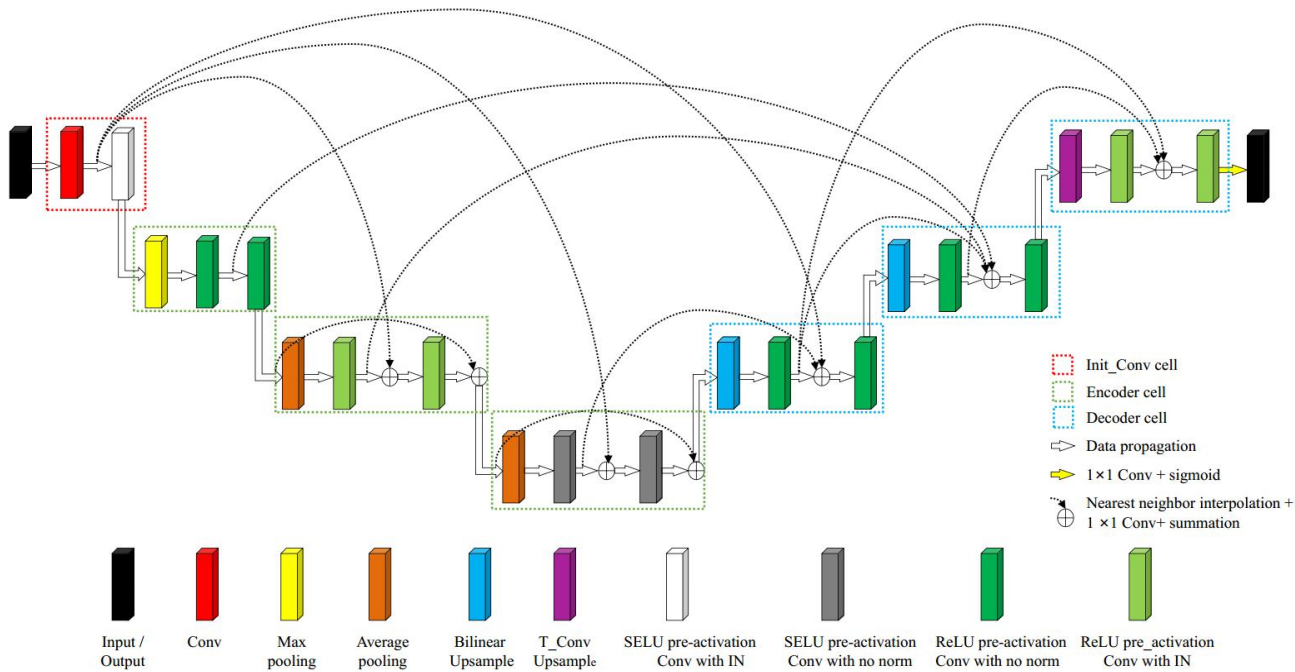

Figure 5. The evolved model by our method on DRIVE dataset

### 5.3 Experiments with the searched model

The evolved model (searched on DRIVE dataset) is evaluated on DRIVE, STARE and CHASE_DB1.

The parameter settings of the algorithm (loss function, optimizer, etc.) are almost the same as those used in training the model during the process of architecture evolution. The only difference is that data augmentation is carried out to avoid overfitting which includes random horizontal and vertical flipping, and random rotation in the range of [-180°,180°]. Unlike the patch-based method, we use the full image as the model input. Due to limitation of the GPU memory (12G), we set the batch size of the experiments as 1 and train the models for 500 epochs.

### 5.3.1 Comparison with existing methods

Table 5, Table 6, and Table 7 show the comparison of our method with the existing methods on the three datasets DRIVE, STARE, CHASE_DB1, respectively. The metric data of the existing methods in these three tables are obtained from the original papers, which include representative methods based on deep learning that have achieved excellent results in retinal vessel segmentation. For DRIVE, our method achieves the best result in four of five metrics. Only SP (specificity) is slightly lower than Fan et al.`s method [21] (0.9835<0.9849). For STARE and CHASE_DB1, our method achieves the best results in all five metrics compared to other methods with a considerable margin. More importantly, for all the three datasets, our method achieves the best results on two comprehensive metrics (F1 score and AUROC), which strongly indicates the superiority of the proposed method.

Table 4. Comparison with existing methods on DRIVE dataset.

| Methods | Year | ACC | SE | SP | Fl | AUROC |
|---|---|---|---|---|---|---|
| Vega *et al.* [20] | 2015 | 0.9412 | 0.7444 | 0.9612 | 0.6884 | N/A |
| Fan *et al.* [21] | 2016 | 0.9614 | 0.7191 | **0.9849** | N/A | N/A |
| Fan and Mo [22] | 2016 | 0.9612 | 0.7814 | 0.9788 | N/A | N/A |
| Liskowski *et al.* [23] | 2016 | 0.9535 | 0.7811 | 0.9807 | N/A | 0.979 |
| Li *et al.* [24] | 2016 | 0.9527 | 0.7569 | 0.9816 | N/A | 0.9738 |
| Orlando *et al.* [25] | 2016 | N/A | 0.7897 | 0.9684 | 0.7857 | N/A |
| Mo and Zhang [26] | 2017 | 0.9521 | 0.7779 | 0.9780 | N/A | 0.9782 |
| Xiao *et al.* [27] | 2018 | 0.9655 | 0.7715 | N/A | N/A | N/A |
| Alom *et al.* [28] | 2019 | 0.9556 | 0.7792 | 0.9813 | 0.8171 | 0.9784 |
| Jin *et al.* [31] | 2019 | 0.9566 | 0.7963 | 0.9800 | 0.8237 | 0.9802 |
| Bo Wang *et al.* [30] | 2019 | 0.9567 | 0.7940 | 0.9816 | 0.8270 | 0.9772 |
| Mou Lei *et al.* [40] | 2019 | 0.9594 | 0.8126 | 0.9788 | N/A | 0.9796 |
| Yicheng Wu *et al.* [29] | 2019 | 0.9578 | 0.8038 | 0.9802 | N/A | 0.9821 |
| The Proposed Method | 2020 | **0.9702** | **0.8341** | 0.9835 | **0.8297** | **0.9882** |

Table 5. Comparison with existing methods on STARE dataset.

| Methods | Year | ACC | SE | SP | Fl | AUROC |
|---|---|---|---|---|---|---|
| Vega *et al.* [20] | 2015 | 0.9483 | 0.7019 | 0.9671 | 0.6614 | N/A |
| Fan *et al.* [21] | 2016 | 0.9588 | 0.6996 | 0.9787 | N/A | N/A |
| Fan and Mo [22] | 2016 | 0.9654 | 0.7834 | 0.9799 | N/A | N/A |
| Liskowski *et al.* [23] | 2016 | 0.9729 | 0.8554 | 0.9862 | N/A | 0.9928 |
| Li *et al.* [24] | 2016 | 0.9628 | 0.7726 | 0.9844 | N/A | 0.9879 |
| Orlando *et al.* [25] | 2017 | N/A | 0.7680 | 0.9738 | 0.7644 | N/A |
| Mo and Zhang [26] | 2018 | 0.9674 | 0.8147 | 0.9844 | N/A | 0.9885 |
| Xiao *et al.* [27] | 2019 | 0.9693 | 0.7469 | N/A | N/A | N/A |

| Alom *et al.* [28] | 2019 | 0.9712 | 0.8292 | 0.9862 | 0.8475 | 0.9914 |
| Jin *et al.* [31] | 2019 | 0.9641 | 0.7595 | 0.9878 | 0.8143 | 0.9832 |
| The Proposed Method | 2020 | **0.9786** | **0.8635** | **0.9881** | **0.8583** | **0.9937** |

Table 6. Comparison with existing methods on CHASE_DB1 dataset.

| Methods | Year | ACC | SE | SP | Fl | AUROC |
| --- | --- | --- | --- | --- | --- | --- |
| Fan and Mo [22] | 2016 | 0.9573 | 0.7656 | 0.9704 | N/A | N/A |
| Liskowski *et al.* [23] | 2016 | 0.9628 | 0.7816 | 0.9836 | N/A | 0.9823 |
| Li *et al.* [24] | 2016 | 0.9527 | 0.7569 | 0.9816 | N/A | 0.9738 |
| Orlando *et al.* [25] | 2016 | N/A | 0.7277 | 0.9712 | 0.7332 | N/A |
| Mo and Zhang [26] | 2017 | 0.9581 | 0.7661 | 0.9793 | N/A | 0.9812 |
| Alom *et al.* [28] | 2019 | 0.9634 | 0.7756 | 0.9820 | 0.7928 | 0.9815 |
| Jin *et al.* [31] | 2019 | 0.9610 | 0.8155 | 0.9752 | 0.7883 | 0.9804 |
| Bo Wang *et al.* [30] | 2019 | 0.9661 | 0.8074 | 0.9821 | 0.8037 | 0.9812 |
| Yicheng Wu *et al.* [29] | 2019 | 0.9661 | 0.8132 | 0.9814 | N/A | 0.9860 |
| The Proposed Method | 2020 | **0.9759** | **0.8546** | **0.9839** | **0.8170** | **0.9909** |

#### 5.3.2 Comparison with U-Net based models

Because the proposed method obtains an optimal model evolved from the U-like architecture, we compare the model with two U-Net based models, the original U-Net [2] and the Attention U-Net [19] in this section.

As Table 7 shows, SP for the evolved model, U-Net and Attention U-Net is 0.9835/0.9841/0.9841 on DRIVE, 0.9835/0.9861/0.9870 on CHASE_DB1 and 0.9881/0.9895/0.9893 on STARE, respectively. ACC for the evolved model, U-Net and Attention U-Net is 0.9759/0.9762/0.9764 on CHASE_DB1. The above-mentioned results show that the metrics of the evolved model is slightly lower than U-Net and Attention U-Net. However, due to the data imbalance of RVS, ACC and SP are not the most proper metrics to evaluate the performance of the models. On the other hand, the evolved model achieves better results than U-Net and Attention U-Net in two other comprehensive metrics of F1 score and AUROC, which are more suitable evaluation metrics in this case.

Moreover, we use ROC curves in Figure 7 and PR curves in Figure 8 to compare the models. The larger area under these two curves directly reflects the better performance of the model. For the ROC curve, the closer it is to the upper left corner, the larger the area under the curve, while for the PR curve, the closer it is to the upper right corner, the larger the area under the curve. It can be observed that whether it is ROC or PR curves, the evolved model has the larger area under the curves among the three models.

We also present some example results in Figure 9. The blue pixels in the images indicate false negative, which is from the vessel regions not detected. As we can see, the evolved model detects vessels more accurately than U-Net and Attention U-Net, either from the overall view or from the locally magnified view. It can be easily noticed that the results of the evolved model have fewer blue pixels. It can be further observed that U-Net and Attention U-Net show their limitations on extracting complicated structural features, such as densely intersected and tiny vessels, while the evolved model can extract them much better.

Table 7. Comparison with U-Net and Attention U-Net on three datasets

| Dataset | Models | ACC | SE | SP | F1 | AUROC |
|---|---|---|---|---|---|---|
| DRIVE | U-Net | 0.9696 | 0.8218 | **0.9841** | 0.8249 | 0.9872 |
| | Attention U-Net | 0.9697 | 0.8229 | **0.9841** | 0.8250 | 0.9875 |
| | Searched model | **0.9702** | **0.8341** | 0.9835 | **0.8297** | **0.9882** |
| CHASE_DB1 | U-Net | **0.9764** | 0.8180 | **0.9870** | 0.8137 | 0.9896 |
| | Attention U-Net | 0.9762 | 0.8295 | 0.9861 | 0.8146 | 0.9891 |
| | Searched model | 0.9759 | **0.8546** | 0.9839 | **0.8170** | **0.9909** |
| STARE | U-Net | 0.9750 | 0.7900 | 0.9893 | 0.8139 | 0.9884 |
| | Attention U-Net | 0.9754 | 0.7918 | **0.9895** | 0.8180 | 0.9868 |
| | Searched model | **0.9786** | **0.8635** | 0.9881 | **0.8583** | **0.9937** |

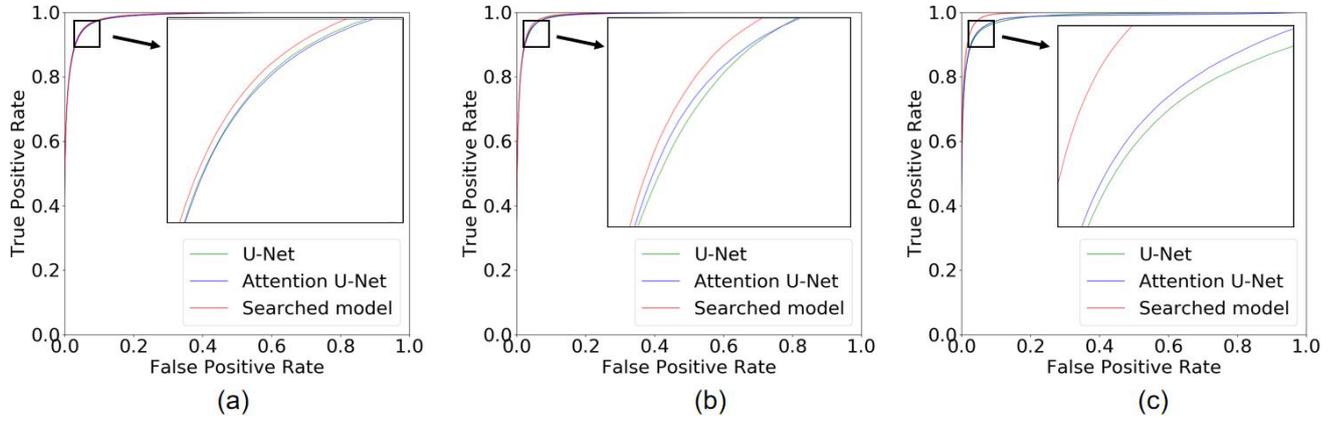

Figure 6. ROC Curves on three datasets. (a) DRIVE. (b) CHASE_DB1. (c) STARE.

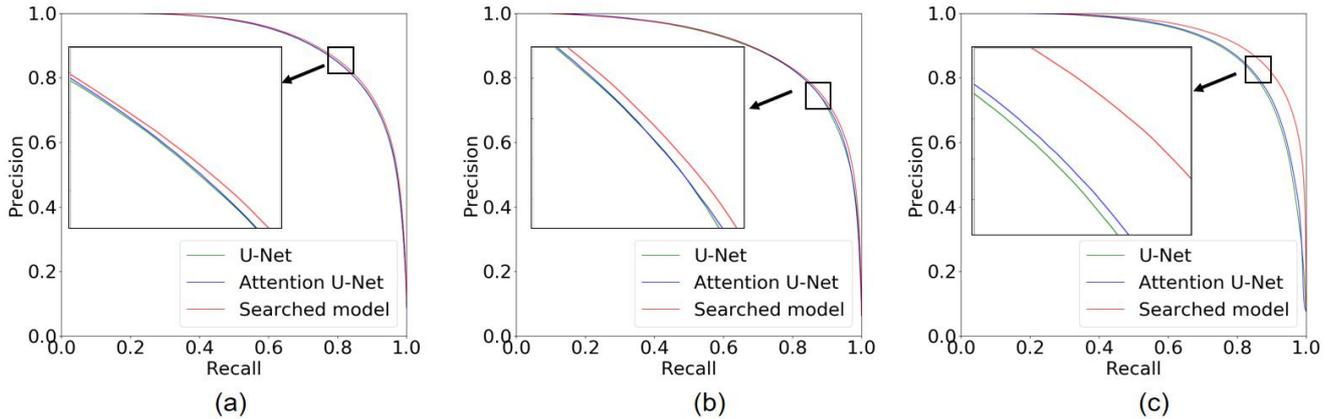

Figure 7. PR Curves on three datasets. (a) DRIVE. (b) CHASE_DB1. (c) STARE.

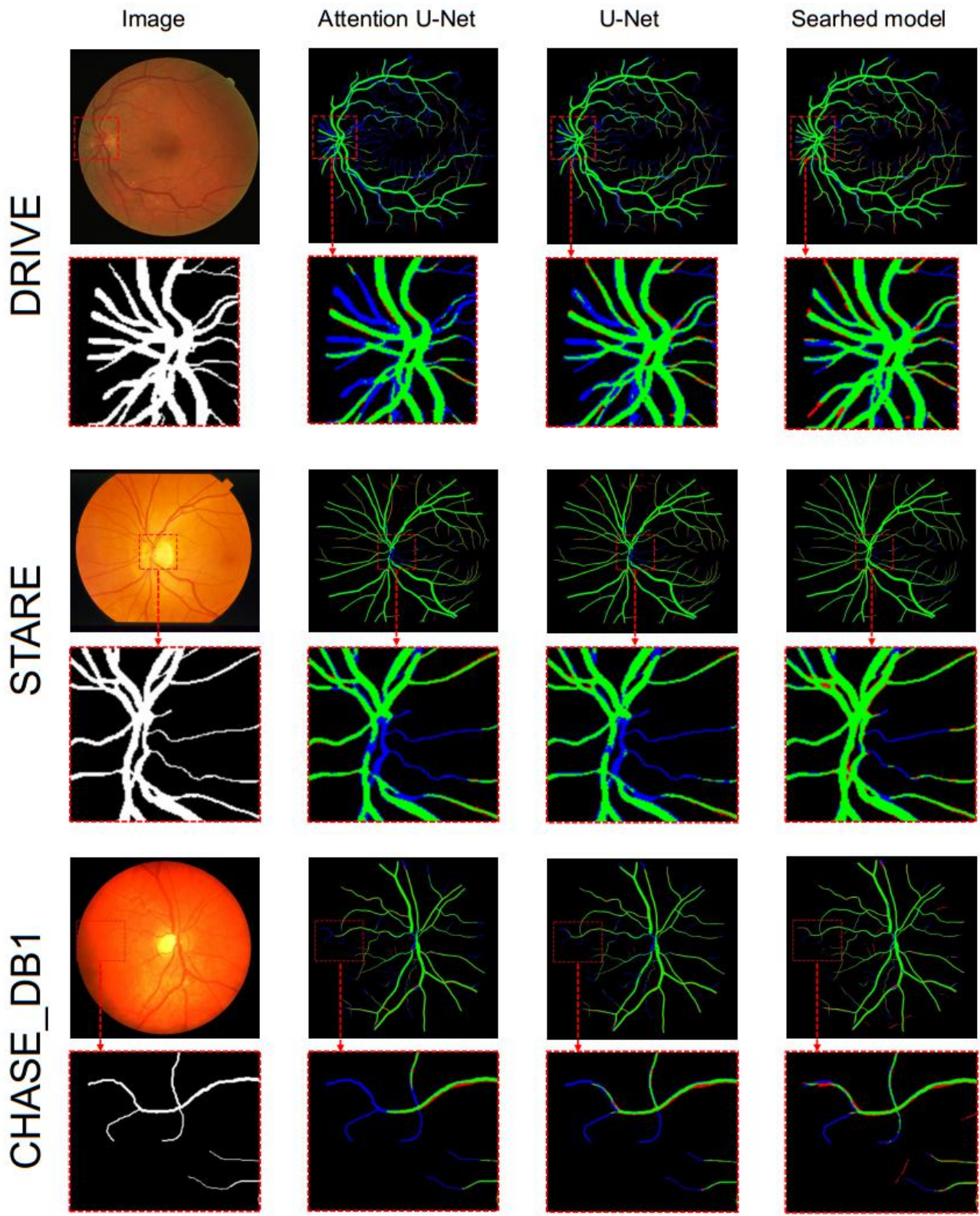

Figure 8. Visualization of the segmentation results on DRIVE, SATRE, and CHASE_DB1 datasets. Green pixel indicates true positive, red pixel indicates false positive and blue pixel indicates false negative.

**5.3.3 Cross-training evaluation**

We also verify the transferability (generalization ability) of the proposed approach by cross-training among the three datasets. The cross-training results of the three models (the evolved model, U-Net and Attention U-net) are presented in Table 9. Compared to training and testing in a single dataset, the performances of the three models are all degraded, because they are trained and tested in two different datasets. However, the evolved model is still the best one, which does not cause a sharp deterioration in performance. After cross-training, the advantages of the evolved model are more prominent because the metric gap between three models is more obvious. One of the most obvious one is that when using STARE as the test set and DRIVE as the training set, the F1 score and AUROC of the evolved model, U-Net and Attention U-Net are 0.5816/0.5646/ 0.7905 and 0.9349/0.9497/0.9829, respectively. When using STARE as the test set and CHASE_DB1 as the training set, the F1 score and AUROC obtained by the evolved model, U-Net and Attention U-Net on STARE are 0.6377 / 0.6142 / 0.7050 and 0.9505 / 0.9203 / 0.9657, respectively. When CHASE_DB1 is used as the test set, a similar situation occurs, and this situation is that U-Net and Attention U-Net perform very poorly with large performance deterioration. A possible explanation for this is that STARE presents more challenges than the other datasets[31][41]. But even in these cases, the evolved model can still obtain satisfactory performance, which indicates that the evolved model can be transferred to new datasets with robust performance, and therefore has strong potential for clinical applications.

Table 8. Comparisons of the cross-training evaluation.

| Dataset | Models/Methods | ACC | SE | SP | F1 | AUROC |
|---|---|---|---|---|---|---|
| DRIVE (trained on STARE) | U-Net | 0.9647 | 0.7303 | 0.9874 | 0.7823 | 0.9749 |
| | Attention U-Net | 0.9663 | 0.7156 | **0.9906** | 0.7868 | 0.9794 |
| | Searched model | **0.9672** | **0.7376** | 0.9896 | **0.7966** | **0.9819** |
| DRIVE (trained on CHASE_DB1) | U-Net | 0.9565 | 0.6064 | 0.9903 | 0.7063 | 0.9568 |
| | Attention U-Net | 0.9565 | 0.5665 | **0.9941** | 0.6922 | **0.9592** |
| | Searched model | **0.9578** | **0.6637** | 0.9863 | **0.7327** | 0.9587 |
| STARE (trained on DRIVE) | U-Net | 0.9487 | 0.5942 | 0.9755 | 0.5816 | 0.9349 |
| | Attention U-Net | 0.9513 | 0.5359 | 0.9831 | 0.5646 | 0.9497 |
| | Searched model | **0.9684** | **0.8065** | 0.9814 | **0.7905** | **0.9829** |
| STARE (trained on CHASE_DB1) | U-Net | 0.9562 | 0.6003 | 0.9839 | 0.6377 | 0.9505 |
| | Attention U-Net | 0.9553 | 0.5676 | **0.9854** | 0.6142 | 0.9203 |
| | Searched model | **0.9567** | **0.7517** | 0.9724 | **0.7050** | **0.9657** |
| CHASE_DB1 (trained on DRIVE) | U-Net | 0.9278 | 0.4704 | 0.9633 | 0.4591 | 0.9240 |
| | Attention U-Net | 0.9407 | 0.3641 | **0.9865** | 0.4421 | 0.9110 |
| | Searched model | **0.9685** | **0.8241** | 0.9801 | **0.7940** | **0.9829** |
| CHASE_DB1 (trained on STARE) | U-Net | 0.9463 | 0.5690 | 0.9745 | 0.5926 | 0.9355 |
| | Attention U-Net | 0.9501 | 0.6070 | **0.9756** | 0.6248 | 0.9304 |
| | Searched model | **0.9567** | **0.7517** | 0.9724 | **0.7050** | **0.9567** |

## 6. Conclusion

In this paper, a novel method of neural architecture search (NAS) is applied to retinal vessel segmentation. Based on the proposed specific search space, we adopt a modified evolutionary strategy to search the macro-architecture of the neural network. Compared to the other models, the evolved model is able to capture more features about the complicated vascular trees from fundus images and provide better segmentation results with fewer parameters. Furthermore, the proposed model can be transferred to new datasets and still perform well in vessel segmentation, which indicates that the model has strong potential for clinical applications. We expect that proposed approach of NAS can be extended to other related applications, such as pavement crack segmentation or semantic segmentation of urban scenes.

# References

[1] Long, Jonathan, Evan Shelhamer, and Trevor Darrell. "Fully convolutional networks for semantic segmentation." Proceedings of the IEEE conference on computer vision and pattern recognition. 2015.

[2] Ronneberger, Olaf, Philipp Fischer, and Thomas Brox. "U-net: Convolutional networks for biomedical image segmentation." International Conference on Medical image computing and computer-assisted intervention. Springer, Cham, 2015.

[3] Gong, Xinyu, et al. "AutoGAN: Neural Architecture Search for Generative Adversarial Networks." Proceedings of the IEEE International Conference on Computer Vision. 2019.

[4] Liu, Chenxi, et al. "Auto-deeplab: Hierarchical neural architecture search for semantic image segmentation." Proceedings of the IEEE Conference on Computer Vision and Pattern Recognition. 2019.

[5] Miikkulainen, Risto, et al. "Evolving deep neural networks." Artificial Intelligence in the Age of Neural Networks and Brain Computing. Academic Press, 2019. 293-312.

[6] Simonyan, Karen, and Andrew Zisserman. "Very deep convolutional networks for large-scale image recognition." ICLR, 2015.

[7] Nair, Vinod, and Geoffrey E. Hinton. "Rectified linear units improve restricted boltzmann machines." Proceedings of the 27th international conference on machine learning (ICML-10). 2010.

[8] Klambauer, Günter, et al. "Self-normalizing neural networks." Advances in neural information processing systems. 2017.

[9] Ioffe, Sergey, and Christian Szegedy. "Batch normalization: Accelerating deep network training by reducing internal covariate shift." arXiv preprint arXiv:1502.03167 (2015).

[10] Ulyanov, Dmitry, Andrea Vedaldi, and Victor Lempitsky. "Instance normalization: The missing ingredient for fast stylization." arXiv preprint arXiv:1607.08022 (2016).

[11] Lu, Zhichao, et al. "NSGA-Net: neural architecture search using multi-objective genetic algorithm." Proceedings of the Genetic and Evolutionary Computation Conference. ACM, 2019.

[12] Xie, Lingxi, and Alan Yuille. "Genetic cnn." Proceedings of the IEEE International Conference on Computer Vision. 2017.

[13] Staal, Joes, et al. "Ridge-based vessel segmentation in color images of the retina." IEEE transactions on medical imaging 23.4 (2004): 501-509.

[14] Fraz, Muhammad Moazam, et al. "An ensemble classification-based approach applied to retinal blood vessel segmentation." IEEE Transactions on Biomedical Engineering 59.9 (2012): 2538-2548.

[15] Orlando, José Ignacio, Elena Prokofyeva, and Matthew B. Blaschko. "A discriminatively trained fully connected conditional random field model for blood vessel segmentation in fundus images." IEEE transactions on Biomedical Engineering 64.1 (2016): 16-27.

[16] Zhang, Michael, et al. "Lookahead Optimizer: k steps forward, 1 step back." Advances in Neural Information Processing Systems. 2019.

[17] Kingma, Diederik P., and J. Ba. "Adam: A Method for Stochastic Optimization." Computer Science (2014).

[18] Lin, Tsung-Yi, et al. "Focal loss for dense object detection." Proceedings of the IEEE international conference on computer vision. 2017.

[19] Oktay, Ozan, et al. "Attention u-net: Learning where to look for the pancreas." MIDL, 2018.

[20] Vega, Roberto, et al. "Retinal vessel extraction using lattice neural networks with dendritic processing." Computers in biology and medicine 58 (2015): 20-30.\

[21] Fan, Zhun, et al. "Automated blood vessel segmentation in fundus image based on integral channel features and random forests." 2016 12th World Congress on Intelligent Control and Automation (WCICA). IEEE, 2016.

[22] Fan, Zhun, and Jia-Jie Mo. "Automated blood vessel segmentation based on de-noising auto-encoder and neural network." 2016 International Conference on Machine Learning and Cybernetics (ICMLC). Vol. 2. IEEE, 2016.

[23] Liskowski, Paweł, and Krzysztof Krawiec. "Segmenting retinal blood vessels with deep neural networks." IEEE transactions on medical imaging 35.11 (2016): 2369-2380.

[24] Li, Qiaoliang, et al. "A cross-modality learning approach for vessel segmentation in retinal images." IEEE transactions on medical imaging 35.1 (2015): 109-118.

[25] Orlando, José Ignacio, Elena Prokofyeva, and Matthew B. Blaschko. "A discriminatively trained fully connected conditional random field model for blood vessel segmentation in fundus images." IEEE transactions on Biomedical Engineering 64.1 (2016): 16-27.

[26] Mo, Juan, and Lei Zhang. "Multi-level deep supervised networks for retinal vessel segmentation." International journal of computer assisted radiology and surgery 12.12 (2017): 2181-2193.

[27] Xiao, Xiao, et al. "Weighted Res-UNet for High-Quality Retina Vessel Segmentation." 2018 9th International Conference on Information Technology in Medicine and Education (ITME). IEEE, 2018.

[28] Alom, Md Zahangir, et al. "Recurrent residual U-Net for medical image segmentation." Journal of Medical Imaging 6.1 (2019): 014006.

[29] Wu, Yicheng, et al. "Vessel-Net: retinal vessel segmentation under multi-path supervision." International Conference on Medical Image Computing and Computer-Assisted Intervention. Springer, Cham, 2019.

[30] Wang, Bo, Shuang Qiu, and Huiguang He. "Dual Encoding U-Net for Retinal Vessel Segmentation." International Conference on Medical Image Computing and Computer-Assisted Intervention. Springer, Cham, 2019.

[31] Jin, Qiangguo, et al. "DUNet: A deformable network for retinal vessel segmentation." Knowledge-Based Systems 178 (2019): 149-162.

[32] Fraz, Muhammad Moazam, et al. "Delineation of blood vessels in pediatric retinal images using decision trees-based ensemble classification." International journal of computer assisted radiology and surgery 9.5 (2014): 795-811.

[33] Fraz, Muhammad Moazam, et al. "An ensemble classification-based approach applied to retinal blood vessel segmentation." IEEE Transactions on Biomedical Engineering 59.9 (2012): 2538-2548.

[34] Zhao, He, et al. "Supervised segmentation of un-annotated retinal fundus images by synthesis." IEEE transactions on medical imaging 38.1 (2018): 46-56.

[35] Meyer, Maria Ines, et al. "A deep neural network for vessel segmentation of scanning laser ophthalmoscopy images." International Conference Image Analysis and Recognition. Springer, Cham, 2017.

[36] Suganuma M, Kobayashi M, Shirakawa S, et al. Evolution of Deep Convolutional Neural Networks Using Cartesian Genetic Programming[J]. Evolutionary computation, 2020, 28(1): 141-163.

[37] He, Kaiming, et al. "Deep residual learning for image recognition." Proceedings of the IEEE conference on computer vision and pattern recognition. 2016.

[38] Huang, Gao, et al. "Densely connected convolutional networks." Proceedings of the IEEE conference on computer vision and pattern recognition. 2017.

[39] Hoover, A. D. , V. Kouznetsova , and M. Goldbaum . "Locating Blood Vessels in Retinal Images by Piecewise Threshold Probing of a Matched Filter Response." IEEE Transactions on Medical Imaging 19.3(2000):203-210.

[40] Mou, Lei, et al. "Dense Dilated Network with Probability Regularized Walk for Vessel Detection." IEEE transactions on medical imaging (2019).

[41] Yan, Zengqiang, Xin Yang, and Kwang-Ting Tim Cheng. "A three-stage deep learning model for accurate retinal vessel segmentation." IEEE journal of biomedical and health informatics (2018).

[42] Bowen Baker, Otkrist Gupta, Nikhil Naik, and Ramesh Raskar. Designing neural network architectures using reinforcement learning. ICLR, 2017.

[43] Barret Zoph and Quoc V Le. Neural architecture search with reinforcement learning. ICLR, 2017.

[44] Barret Zoph, Vijay Vasudevan, Jonathon Shlens, and Quoc V Le. Learning transferable architectures for scalable image recognition. In CVPR, pages 8697–8710, 2018.

[45] Hanxiao Liu, Karen Simonyan, and Yiming Yang. Darts: Differentiable architecture search. ICLR, 2019.

[46] Andrew Brock, Theodore Lim, James M Ritchie, and Nick Weston. Smash: one-shot model architecture search through hypernetworks. ICLR, 2018.

[47] Ghiasi, Golnaz, Tsung-Yi Lin, and Quoc V. Le. "Nas-fpn: Learning scalable feature pyramid architecture for object detection." Proceedings of the IEEE Conference on Computer Vision and Pattern Recognition. 2019.

[48] Xu, Hang, et al. "Auto-fpn: Automatic network architecture adaptation for object detection beyond classification." Proceedings of the IEEE International Conference on Computer Vision. 2019.

[49] Zhuotun Zhu, Chenxi Liu, Dong Yang, Alan Yuille, and Daguang Xu. V-nas: Neural architecture search for volumetric medical image segmentation. 3DV, 2019.

[50] Dong Yang, Holger Roth, Ziyue Xu, Fausto Milletari, Ling Zhang, and Daguang Xu. Searching learning strategy with reinforcement learning for 3d medical image segmentation. In MICCAI, pages 3–11. Springer, 2019.

[51] Aliasghar Mortazi and Ulas Bagci. Automatically designing cnn architectures for medical image segmentation. In MLMI, pages 98–106. Springer, 2018.

[52] Yu Weng, Tianbao Zhou, Yujie Li, and Xiaoyu Qiu. Nas-unet: Neural architecture search for medical image segmentation. IEEE Access, 7:44247–44257, 2019.

[53] Sungwoong Kim, Ildoo Kim, Sungbin Lim, Woonhyuk Baek, Chiheon Kim, Hyungjoo Cho, Boogeon Yoon, and Taesup Kim. Scalable neural architecture search for 3d medical image segmentation. arXiv preprint arXiv:1906.05956, 2019.

[54] Laibacher, Tim, Tillman Weyde, and Sepehr Jalali. "M2U-Net: Effective and Efficient Retinal Vessel Segmentation for Real-World Applications." Proceedings of the IEEE Conference on Computer Vision and Pattern Recognition Workshops. 2019.

[55] Yan, Zengqiang, Xin Yang, and Kwang-Ting Cheng. "Joint segment-level and pixel-wise losses for deep learning based retinal vessel segmentation." IEEE Transactions on Biomedical Engineering 65.9 (2018): 1912-1923.

[56] Chatziralli, Irini P., et al. "The value of fundoscopy in general practice." The open ophthalmology journal 6 (2012): 4.

[57] Fraz, M. M. , et al. "Blood vessel segmentation methodologies in retinal images – A survey." Computer Methods and Programs in Biomedicine 108.1(2012):407-433.

[58] Vostatek, Pavel , et al. "Performance Comparison of Publicly Available Retinal Blood Vessel Segmentation Methods." Computerized Medical Imaging and Graphics 55(2016):2-12.

[59] Ortega, Marcos , et al. "Personal verification based on extraction and characterization of retinal feature points." Journal of Visual Languages & Computing 20.2(2009):80-90.

[60] Simon and I. Goldstein. A new scientific method of identification. New York State Journal of Medicine, 35(18):901–906, Sept. 1935

[61] Sandler, Mark, et al. "Mobilenetv2: Inverted residuals and linear bottlenecks." Proceedings of the IEEE Conference on Computer Vision and Pattern Recognition. 2018.

[62] Beyer, Hans-Georg, and Hans-Paul Schwefel. "Evolution strategies–A comprehensive introduction." Natural computing 1.1 (2002): 3-52.